# A Bayesian Bootstrap Approach for Dynamic Borrowing for Minimizing Mean Squared Error

Jixian Wang[a] and Ram Tiwari[b]

[a]Bristol Myers Squibb, Boudry, Switzerland;
[b]Bristol Myers Squibb, Berkeley Heights, New Jersey, USA.

August 1, 2024

### Abstract

For dynamic borrowing to leverage external data to augment the control arm of small RCTs, the key step is determining the amount of borrowing based on the similarity of the outcomes in the controls from the trial and the external data sources. A simple approach for this task uses the empirical Bayesian approach, which maximizes the marginal likelihood (maxML) of the amount of borrowing, while a likelihood-independent alternative minimizes the mean squared error (minMSE). We consider two minMSE approaches that differ from each other in the way of estimating the parameters in the minMSE rule. The classical one adjusts for bias due to sample variance, which in some situations is equivalent to the maxML rule. We propose a simplified alternative without the variance adjustment, which has asymptotic properties partially similar to the maxML rule, leading to no borrowing if means of control outcomes from the two data sources are different and may have less bias than that of the maxML rule. Consequently, it may not lead to full borrowing, even if the two data sources are identical. We argue that this may not be a practical problem, since more than often a cap of the amount of external data to be borrowed is applied anyway. In contrast, the maxML rule may lead to full borrowing even when two datasets are moderately different, which may not be a desirable property. For inference, we propose a Bayesian bootstrap (BB) based approach taking the uncertainty of the estimated amount of borrowing and that of pre-adjustment into account. The approach can also be used with a pre-adjustment on the external controls for population difference between the two data sources using, e.g., inverse probability weighting. The proposed approach is computationally efficient and is implemented via a simple algorithm. We conducted a simulation study to examine properties of the proposed approach, including the coverage of 95 percent confidence interval based on the Bayesian bootstrapped posterior samples, or asymptotic normality. The approach is illustrated by an example of borrowing controls for an AML trial from another study.

## 1 Introduction

When a randomized clinical trial (RCT) has a small control arm, external data, which may come from another clinical trial or real-world data (RWD), may be a valuable source for augmenting the control of the RCT. This approach is often referred to as borrowing external controls. Regulatory guidance documents on the use of RWD to aid drug development have been published [1, 2]. However, the use of external data also introduces the risk of potential bias, as the historical control population may be rather different from the RCT. To mitigate the risk, the control arm in an RCT, may provide an internal reference for evaluating the difference in the outcomes between the internal and external control data sources. To address the population difference, the power prior approach [3, 4, 5, 6, 7, 8, 9, 10] can be used to discount historical data to mitigate the impact of bias. The amount of borrowing, parameterized by the power prior parameter, can be fixed or determined based on the similarity of the outcomes or key prognostic factors of the two data sources, known as Bayesian dynamic borrowing.

A simple dynamic borrowing approach based on the empirical Bayes approach determines the amount of borrowing by maximizing marginal likelihood of the power prior parameter based on normal or binomial likelihood functions [11, 12]. This approach will be referred to as the maxML approach, henceforth. The combined estimator of the mean outcome under the augmented control is consistent even when the means of the outcomes in the internal and external controls are different [13].

The difference in outcomes may be partly due to the difference in baseline covariates between the two data sources. Hence, approaches such as propensity score [14] based weighting, known as inverse probability weighting (IPW), matching or stratification can be used to reduce or eliminate the difference [15, 16, 9]. An alternative approach uses a linear outcome model that assumes exchangeability after covariate adjustment [17]. However, often unobserved confounding factors exist; hence a considerable difference may still exist after adjustment and lead to a moderate bias. Therefore, the dynamic power approach should still be used as a safeguard after adjustment. To take into account variability in both adjustment and determination of the amount of borrowing in a full Bayesian analysis, Wang et al. [13] proposed a Bayesian bootstrap (BB) based approach [18] combining the maxML and IPW adjustment for approximate Bayesian inference.

As an alternative to maxML, we propose using likelihood-independent rules to determine the amount of borrowing by minimizing the posterior mean squared error (MSE) of the combined estimator, which will be referred to as the minMSE approach hereafter. In fact, approaches for minimizing MSE has a long history; see Galwey [19] for previous works and his numerical results in a different setting, in which he showed that dynamic borrowing had minimum benefit even in the best scenarios. The minMSE rule depends on unknown parameters to be estimated from the data. The classical minMSE approach (cminMSE hereafter) plugs in an unbiased estimate for the squared mean difference to adjust for the variance component in it. In the scenario of normal distributed response, the cminMSE and maxML rules are equivalent. As an alternative, we propose a simple version which uses the squared sample mean difference without variance adjustment. Consequently, this version of minMSE generally borrows less than the other version and may not borrow in full when the internal and external data sources are identical. Nevertheless, we argue that this may not be a serious problem in practice, since often we need a cap on the amount of borrowing to mitigate the impact of external data.

For statistical inference, we propose a BB approach taking into account the uncertainty of the estimated amount of borrowing. The approach does not depend on a full likelihood specification, and hence is simple and flexible [20, 21]. The simple minMSE rule is a continuous function of the sample mean and variance with bounded derivatives, hence the asymptotic properties of the BB estimator for treatment effect can be easily derived. For small sample properties, our simulation shows that the minMSE approach may result in significantly lower MSE than maxML or cminMSE, except when the two data sources are very similar.

The paper is organized as follows. The next section introduces the framework of Bayesian borrowing for RCT. Section 3 briefly reviews Bayesian borrowing approaches, in particular the maxML approach. In the following section, we propose our minMSE approach that can be adapted for estimating different estimands using outcomes from different outcome distributions. Section 5 describes the BB approach and a detailed algorithm for implementing it in minMSE. A simulation study is reported in Section 6, followed by an application to illustrate the use of the minMSE approach to acute myeloid leukemia (AML) trials in Section 7. We end with a discussion of the extension of our approach to include, e.g., covariate adjustment using different approaches.

## 2  Causal treatment estimand with borrowing external controls

Let $Y_{ij}$ and $Z_{ij}$ denote the outcome and treatment indicator for subject $i$ of population $j$, $j = 1$ for the external data source; and $j = 0$ otherwise. There are $n_j$ subjects in the population $j$. As an indicator, let $Z_{ij} = t$ denote the test treatment and $Z_{ij} = c$ denote the control. In our situation, we have $Z_{i1} = c$; that is, the external source is only for control. In addition, we use $D_j$ to denote the dataset in population $j$, $j = 0, 1$.

Our final estimand of interest is the causal treatment effect of the test treatment, compared with the control. Formally, we take the Bayesian causal framework [22, 23], and are interested in the causal effect defined as

$$\tau = E(Y_{i0}(t)) - E(Y_{i0}(c)) \equiv \mu^t - \mu^c, \tag{1}$$

where $Y_{ij}(t)$ and $Y_{ij}(c)$ are the (counterfactual) outcomes of subject $i$ from population $j$ when treated with $t$ and $c$, respectively. The expectation is taken over the super trial population, not the finite population of the $n_0$ subjects, but we will drop "superscript" hereafter for simplicity. The above estimand is also known as the average treatment effect of the treated (ATT), since the trial population is the target population for the test treatment. Although our final goal is inference for $\tau$, as inference for $\mu^t$ can easily use data from the treated arm in the trial, the challenge is inference for $\mu^c$ using control subjects in both $D_0$ and $D_1$, which will be the focus of this paper. As we will not deal with the test arm, to simplify the notation, we ignore this arm and assume that all $n_0$ subjects are in the control arm. For later use, we denote the means of control patients from populations $j = 1$ and $j = 0$ by $\mu_1 = E(Y_{i1}(c))$ and $\mu_0 = E(Y_{i0}(c))$, respectively. Due to randomization in the trial, we have $\mu^c = \mu_0$ and will use notation $\mu_0$ rather than $\mu_c$, hereafter. Specifically, our goal is to use data $D_0, D_1$ for statistical inference regarding $\mu_0$, or equivalently, $\mu_c$.

## 3  Power prior and Bayesian dynamic borrowing

For borrowing data $D_1$ to improve the inference of $\mu_0$, Bayesian borrowing with the power prior is a powerful tool [4, 3, 5, 6, 24, 11, 12, 15]. First, we introduce this approach in a general form and then describe the specific one for the inference of $\mu_c$. The approach is based on the likelihood function with parameters $\theta$. Let $L(\theta|D)$ denote the likelihood function, given data $D$. The power prior, conditional on $D_1$, is formulated as

$$\pi(\theta|D_1, a_0) \propto L(\theta|D_1)^{a_0} \pi_0(\theta), \tag{2}$$

where $0 \leq a_0 \leq 1$ is the power prior (discounting) parameter in the likelihood of external data, and $\pi_0(\theta)$ is the initial prior for $\theta$. The corresponding posterior distribution is

$$\pi(\theta|D_1, D_0, a_0) \propto L(\theta|D_0) L(\theta|D_1)^{a_0} \pi_0(\theta). \tag{3}$$

The power prior assumes the same parameters $\theta$ for $D_1$ and $D_0$. The commensurate power priors contains separate parameters $\theta_1$ and $\theta_0$ for them:

$$\pi(\theta_0, \theta_1 | D_1, a_0) \propto L(\theta_1 | D_1)^{a_0} \pi_0(\theta | \theta_1) \pi_0(\theta_1), \tag{4}$$

In the conditional prior $\pi_0(\theta | \theta_1)$ one can introduce a parameter to represent commensurability between $\theta_0$ and $\theta_1$.

With a fixed $a_0$, normally distributed outcomes $Y_{i0}$ and $Y_{i1}$, with common variance $\sigma^2$ respectively, and the prior assumptions as detailed in [11], the posterior distribution of $\mu_0$ is

$$\mu_0 \sim N(\hat{\mu}(a_0), \hat{\sigma}^2) \tag{5}$$

where

$$\hat{\mu}_0(a_0) = \hat{\sigma}^2 (a_0 \sigma_1^{-2} \bar{y}_1 + \sigma_0^{-2} \bar{y}_0) \tag{6}$$

$$\hat{\sigma}^2 = (a_0 \sigma_1^{-2} + \sigma_0^{-2})^{-1} \tag{7}$$

and $\bar{y}_j$ is the sample means of $Y_{ij}$ with variances $\sigma_j^2 = \sigma_e^2 / n_j$, where $\sigma_e^2 = \text{var}(Y_{ij})$ is known. In practice, the variances can be replaced by the sample variances. Using a location-commensurate power prior [5, 24] with $\pi_0(\mu_0 | \mu_1) \sim N(\mu_1, \eta)$ in our situation, the posterior mean of $\mu_0$ is also a weighted mean of $\bar{y}_1$ and $\bar{y}_0$ (Section 2.3, Ref [5]).

For binomial outcomes, with a $Beta(1,1)$ prior and a fixed $a_0$, the posterior distribution of $\mu_c$ can be written as

$$\mu_0 \sim Beta(a_0 y_{1.} + y_{0.} + 1, n_0 + a_0(n_1 - y_{1.}) - y_{0.} + 1), \tag{8}$$

where $y_{1.} = \sum_{i=1}^{n_1} y_{i1}$ and $y_{0.} = \sum_{i=1}^{n_0} y_{i0}$ [11].

To determine $a_0$ based on the data, maxML is a simple approach [11, 12]. This approach estimates $a_0$ by maximizing the marginal likelihood for $a_0$; $\pi(a_0 | D_1, D_0)$. For the normal distribution case above, such an $a_0$ has a closed form:

$$\hat{a}_0 = \frac{\sigma_1^2}{\max[(\bar{y}_1 - \bar{y}_0)^2, \hat{\sigma}_1^2 + \hat{\sigma}_0^2] - \hat{\sigma}_0^2}. \tag{9}$$

Then, we can estimate the mean of $\mu_c$ by plugging $\hat{a}_0$ into (6) and (7).

The same approach can be used for binary outcomes. The marginal likelihood of $a_0$ is (Eq 4 of Ref [11])

$$\pi(a_0 | D_1, D_0) \propto \frac{Beta(a_0 y_{1.} + y_{0.} + 1, a_0(n_1 - y_{1.}) + n_0 - y_{0.} + 1)}{Beta(a_0 y_{1.} + 1, a_0(n_1 - y_{1.}) + 1)}. \tag{10}$$

Gravestock & Held [11] proposed to use

$$\hat{a}_0 = \text{argmax}_{a_0} \pi(a_0 | D_1, D_0) \tag{11}$$

within the range of [0, 1]. In practice, $\hat{a}_0$ can be found with a grid search.

## 4 Borrowing to minimize MSE

Here, we consider how to combine the two data sources and determine the amount of borrowing to minimize the MSE of a linear combination estimator such as (6). This part is not Bayesian; hence we assume known parameter values to derive the amount of borrowing, then replace the unknown parameters with their estimates. Specifically, we consider a linear combination of $\hat{\mu}_1 \equiv \bar{y}_1$ and $\hat{\mu}_0 \equiv \bar{y}_0$ as an estimator of $\mu_c$:

$$\hat{\mu}_c(a) = \frac{\hat{\mu}_0 + a\hat{\mu}_1}{1+a}. \tag{12}$$

This general form includes that of maxML (6) in the normal case with $a = a_0 \sigma_0^2/\sigma_1^2 = a_0 n_1/n_0$. Therefore, for full borrowing, i.e., $a_0 = 1$, we need $a = n_1/n_0$. We want to find the $a$ that minimizes the posterior MSE of $\hat{\mu}_c(a)$, that is,

$$a^* = \mathrm{argmin}_a MSE(\hat{\mu}_c(a)), \tag{13}$$

where

$$MSE(\mu_c(a)) = \mathrm{var}(\hat{\mu}_c(a)) + \Delta^2(a) \tag{14}$$

and $\Delta(a) = E(\hat{\mu}_c(a)) - \mu_0$ is the bias of $\hat{\mu}_c(a)$. Denoting the mean difference by $\delta = E(\hat{\mu}_1) - E(\hat{\mu}_0) = \mu_1 - \mu_0$, with some simple derivation, the variance and bias of $\hat{\mu}_c(a)$ can be written as

$$\mathrm{var}(\hat{\mu}^c(a)) = \frac{\sigma_0^2 + a^2 \sigma_1^2}{(1+a)^2}; \quad \Delta(a) = \frac{a\delta}{1+a}. \tag{15}$$

These lead to

$$MSE(\hat{\mu}^c(a)) = \frac{\sigma_0^2 + a^2(\sigma_1^2 + \delta^2)}{(1+a)^2}. \tag{16}$$

and the $a$ that minimizes it is

$$a^* = \sigma_0^2/(\sigma_1^2 + \delta^2). \tag{17}$$

With this $a^*$, the bias can derived as

$$\Delta(a^*) = \frac{\delta \sigma_0^2}{\sigma_1^2 + \sigma_0^2 + \delta^2}. \tag{18}$$

The Appendix shows that the maximum squared bias $\Delta^2(a^*)$ occurs when $\delta^2 = \sigma_1^2 + \sigma_0^2$.

To use the minMSE estimators, we need to replace the variances and bias by their estimates. The cminMSE rule, which is examined in details in Galwey [19], but proposed by others much earlier, replaces $\delta^2$ by its unbiased estimate $\hat{\delta}^2 - \hat{\sigma}_1^2 - \hat{\sigma}_0^2$, leading to the cminMSE rule

$$\hat{a}_c^* = \hat{\sigma}_0^2/[\max(\hat{\delta}^2, \hat{\sigma}_1^2 + \hat{\sigma}_0^2) - \hat{\sigma}_0^2] = \hat{\sigma}_0^2/\max(\hat{\delta}^2 - \hat{\sigma}_0^2, \hat{\sigma}_1^2). \tag{19}$$

Interestingly, it is equivalent to $\hat{a}_0$ in (9), considering the relationship $a = a_0 \sigma_0^2/\sigma_1^2$.

The cminMSE approach may have $\hat{a}_c^* = 1$ for moderate $\hat{\delta}^2$ relative to the total variance, which may lead to considerable bias. To avoid this issue, we propose a our minMSE rule in which $\delta^2$ is replaced by $\hat{\delta}^2$ without the variance correction, i.e.

$$\hat{a}^* = \hat{\sigma}_0^2/(\hat{\sigma}_1^2 + \hat{\delta}^2) \tag{20}$$

Note that (17) is derived based on known parameter values, and hence neither the cminMSE nor minMSE rules truly minimize the MSE.

Although both $\hat{a}_0$ and $\hat{a}^*$ have no upper limit, in practice, it is advisable to set such a limit to control the impact of external data on $\hat{\mu}_c(a)$.

Next, we consider the asymptotic properties of $\hat{a}^*$ in (20) and $\hat{a}_c^*$, which is equivalent to $\hat{a}_0$, hence we can borrow the properties of the latter. Reference [25] proposed a criterion "information borrowing consistent", which applies to our context is that $\hat{a}^* \to 1$ in probability when $\mu_1 = \mu_0$, and $\hat{a}^* \to 0$ when $\mu_1 \neq \mu_0$ when $n_j \to \infty$. Without loss of generality, we assume that the regularity conditions on $Y_{ij}$ lead to $\hat{\delta} \to \delta$, a.s., and $\text{var}(\bar{y}_{j.}) = \sigma_j^2 = \sigma^2/n_j$. When $\mu_1 \neq \mu_0$, we have $\hat{\delta}^2 \to \delta^2 \neq 0$. Taking it into (20) and (19), we have $\hat{a}^* \to 0$ and $\hat{a}_c^* \to 0$, a.s., hence this part of the criterion is satisfied by both minMSE rules. When $\mu_1 = \mu_0$, since $\hat{\delta}^2 \to \sigma_1^2 + \sigma_0^2$, we have $\max(\hat{\delta}^2 - \hat{\sigma}_0^2, \hat{\sigma}_1^2) \to \sigma_1^2$, then $\hat{a}_c^* \to n_1/n_0$, which leads to full borrowing, if there is no cap on it. However, $\hat{a}^* \to \sigma_0^2/(2\sigma_1^2 + \sigma_0^2) = n_0^{-1}/(2n_1^{-1} + n_0^{-1}) < n_1/n_0$, a.s., and therefore $\hat{a}^*$ does not lead to full borrowing. Although $\hat{a}^*$ has the inconsistency issue in this ideal situation, it may not be a serious practical problem, since the amount of borrowing is often capped and full borrowing, which treats external data the same as internal ones, is often not desirable. Our simulation evidence shows (Section 6) that for a wide range of parameter settings, $\hat{a}^*$ results in a much lower MSE than $\hat{a}_0$ (or equivalently $\hat{a}_c^*$).

The minMSE rule works for a wide range of outcomes, including normal and binary outcomes, since we only need to replace the mean and variance by their estimators in (20). For example, for binary outcomes, we can use $\hat{\mu}_j = y_{j.}/n_j$ and $\hat{\sigma}_j^2 = y_{j.}/n_j(1 - y_{j.}/n_j)$. If appropriate, one may also use the cminMSE rule and the variances in $\hat{\delta}^2$. Finally, although we derived the minMSE rule as a frequentist approach, one can also replace $\delta$ and $\mu_j$ with their Bayesian estimates.

## 4.1 Extensions of the minMSE approach

Although MSE is a good summary taking both bias and variability into account, people may want case-specific relative weights between them. The first extension of our approach is to minimize a weighted MSE

$$MSE(\hat{\mu}_c(a), \eta) = var(\hat{\mu}_c(a)) + \eta^2 \Delta^2(a), \tag{21}$$

where weight $\eta$ represents the relative importance of the bias. The previous results can be easily adapted. Since

$$\Delta(a) = \frac{a\delta}{1+a}, \tag{22}$$

we can write

$$MSE(\hat{\mu}_c(a), \eta) = var(\hat{\mu}_c) + \eta^2 \Delta^2(a), \tag{23}$$

$$= \frac{\sigma_0^2 + a^2(\sigma_1^2 + (\eta\delta)^2)}{(1+a)^2}, \tag{24}$$

which is the same as (16), except that $\delta$ is replaced with $\eta\delta$. Hence, to use the previous results, we only need to replace $\delta$ with $\eta\delta$ in the other formulae.

In the minMSE estimator, we have used data $D_1$ to calculate $\hat{\mu}_1$ directly. When we observe $p$ key prognostic factors $\mathbf{X}_{ij}$ and that are imbalanced between the two populations, adjustment can be carried out for subjects in $D_1$ so that $\mu_1$ becomes more similar to $\mu_0$ by eliminating the impact of difference in $\mathbf{X}_{ij}$ between $j = 1$ and $j = 0$. We briefly describe the IPW approach here, although other adjustment methods can also be used. The IPW approach [14] is based on the propensity score (PS) defined as

$$e_i = P(j = 0|\mathbf{X}_{ij}), \tag{25}$$

which can be estimated by a logistic regression to obtain $\hat{e}_i$ for all subjects in $D_1$. Then an IPW

adjusted estimator is

$$\hat{\mu}_1 = n_1^{-1} \sum_{i=1}^{n_1} \frac{\hat{e}_i}{1-\hat{e}_i} Y_{i1}. \qquad (26)$$

Capistran et al. [26] and Stephens et al. [27] showed how to obtain posterior BB samples using the IPW estimator and that $\hat{\mu}_1$ converges to the posterior mean. Even when the PS model is misspecified, an IPW adjustment may still reduce the difference between $\hat{\mu}_0$ and $\hat{\mu}_1$ and results in more borrowing and smaller MSE than that without adjustment.

## 5 Bayesian bootstrap for approximate Bayesian inference

Here we propose using a BB approach for approximate Bayesian inference based on the estimator $\hat{\mu}_c(\hat{a}^*)$. Since $\hat{\mu}_c(\hat{a}^*)$ contains multiple sources of variability, including those for estimating $a^*$ and adjusting for $X_{ij}$. Consequently, even if we are willing to assume the asymptotic normality of $\hat{\mu}_c(\hat{a}^*)$, calculating its asymptotic variance is not easy. BB provides a simple approach to approximate Bayesian inference by constructing a posterior sample of $\mu_c$. By "approximate" we mean that the BB approach cannot rely on small sample properties that some classical Bayesian methods enjoy when correct priors and models are specified.

Suppose that, based on the data $D_0$ and $D_1$, we can obtain BB posterior samples $\mu_0^b \sim F(\mu|D_0)$ and $\mu_1^b \sim F(\mu|D_1)$, respectively. In particular, we have posterior means and variances, $E(\mu_0^b) = \hat{\mu}_0, E(\mu_1^b) = \hat{\mu}_1$ and $\text{var}(\mu_0^b) = \hat{\sigma}_0^2, \text{var}(\mu_1^b) = \hat{\sigma}_1^2$, which can replace the unknown parameters in (20). Here, we show that BB can be used to obtain posterior samples of $\mu_c$ for statistical inference.

To this end, the first step is to obtain posterior samples of $\mu_1^b$ and $\mu_0^b$ using BB. We draw weights from a non-informative Dirichlet distribution $w_{ij}^b \sim Dir(1,...,1)$ for all subjects in $D_j, j = 0, 1$; b=1,...,B, respectively. Then BB posterior samples for $\mu_j$ and $\sigma_j^2$ can be obtained as the weighted mean

$$\hat{\mu}_j^b = \sum_{i=1}^{n_j} w_{ij}^b Y_{ij}/n_j,$$
$$\hat{\sigma}_j^{2b} = \sum_{i=1}^{n_j} w_{ij}^b (Y_{ij} - \hat{\mu}_j^b)^2/(n_j-1) \quad j=0,1; \ b=1,...,B. \qquad (27)$$

Here, we assume that the weights $w_{ij}^b$ have been normalized so that their sum is $n_j$ rather than 1. We obtain the posterior samples $\hat{\mu}_j^1, ..., \hat{\mu}_j^B, b = 1, ..., B, j = 0, 1$. Consequently, we can construct posterior samples of $\delta$ as $\hat{\delta}^b = \hat{\mu}_1^b - \hat{\mu}_0^b$ and that of $a^*$ by plugging $\hat{\mu}_j^b$s into (20) to obtain $\hat{a}^{*b}$. Taking $\hat{a}^{*b}$, $\hat{\mu}_j^b$ and $\hat{\sigma}_j^{2(b)}$ to replace the sample means and variances in (12), we obtain the posterior samples $\hat{\mu}_c^b(\hat{a}^{*b})$ as desired. Then statistical inference can be based on an approximate normal distribution using posterior mean and variance, or on the empirical distribution of posterior samples $\hat{\mu}_c^b$.

Although the posterior sample $\hat{\mu}_c^b$ can be used for inference without relying on the posterior distribution, it is a theoretical justification to examine its asymptotic distribution, since the minMSE rules do not depend on the full model specification. The asymptotic properties of the BB sample can be found in the appendix together with justification of type I error control.

For the estimation and inference of treatment effect $\tau$ based on BB, we also need to apply the BB step to treated subjects in the trial to obtain $\hat{\mu}_t^b$. Then, one can obtain the posterior samples of $\tau$ by $\hat{\tau}^b = \hat{\mu}_t^b - \hat{\mu}_c^b, b = 1, ..., B$. Algorithm 1 details all the steps, including possible pre-adjustment of $D_1$ data with IPW.

---

**Algorithm 1** Bayesian bootstrap for posterior sample of $\mu_c$.

---
**Require:** $Y_{ij}, \mathbf{X}_{ij}, B$
1: **for** Bootstrap run $b = 1$ to $B$ **do**
2:    Generate weights $w_{ij}^b, i = 1, ..., n_j, j = 0, 1$ from the uniform Dirichlet distribution $Dir(1, ..., 1)$
3:    (Optional for IPW adjustment) Fit a weighted PS model for $j = 0$ including $\mathbf{X}_{ij}$ to obtain $\hat{e}_i^b$.
4:    Obtain

$$\hat{\mu}_j^b = \sum_{i=1}^{n_j} w_{ij}^b Y_{ij}/n_j,$$
$$\hat{\sigma}_j^{2b} = \sum_{i=1}^{n_j} w_{ij}^b (Y_{ij} - \hat{\mu}_j^b)^2/(n_j - 1) \quad j = 0, 1; \; b = 1, ..., B. \quad (28)$$

    or, if use IPW adjustment for $\hat{\mu}_1$, we replace the weights by $\xi_i^b = w_{i1}^b \hat{e}_i^b/(1 - \hat{e}_i^b)$, which is also normalized such that $\sum_{i=1}^{n_1} \xi_i^b = n_1$.

5:    Obtain BB sample variances $\hat{\sigma}_0^{2b}$ and $\hat{\sigma}_1^{2b}$ using the same BB weights.
6:    Obtain $\delta^b = \hat{\mu}_1^b - \hat{\mu}_0^b$ and

$$\hat{a}^{*b} = \hat{\sigma}_0^{2(b)}/(\hat{\sigma}_1^{2(b)} + (\hat{\delta}^b)^2) \quad (29)$$

7:    Calculate

$$\hat{\mu}_c^b = \frac{\hat{\mu}_0^b + \hat{a}^{*b}\hat{\mu}_1^b}{1 + \hat{a}^{*b}} \quad (30)$$

8:    Generate $\hat{\mu}_t^b, b = 1, ..., B$ in the same way as in steps 2 and 4, using the active arm in the trial.
9:    Calculate $\hat{\tau}^b = \hat{\mu}_t^b - \hat{\mu}_c^b$
10: **end for**
11: Output $\hat{\tau}^1, ..., \hat{\tau}^B$.

---

## 6 A simulation study

A simulation study is conducted to examine the performance of the minMSE approach compared to the maxML. Although minMSE can be extended to include pre-adjustment or using weighted MSE, the simulation focuses on the basic approach as this setting is sufficient to capture key features of the minMSE approach. For example, if IPW pre-adjustment reduced $\delta$ from 0.3 to 0.2, then the simulation with $\delta = 0.2$ and $\delta = 0.3$ can reflect the impact of the adjustment. In addition, we concentrate on the property of $\hat{\mu}_c$ rather than $\hat{\tau}$ since the borrowing approach affects the latter only via the former.

The following gives the parameters of the simulation setting.

- As continuous outcomes, we generate $Y_{ij} \sim N(\mu_{ij}, 1)$ or $Y_{ij} \sim t(\mu_{ij}, df)$, with $\mu_{ij} = \mathbf{X}_{ij}^T \boldsymbol{\beta} + j\Delta_x$, where $df = 3$, $\boldsymbol{\beta} = 0.5\,\mathbf{1}$, $\mathbf{1}$ is vector of $p$ ones and $\mathbf{X}_{ij} \sim N(0, 1)$ IID. With this setting, $\Delta_x$ represents the mean difference $\delta$, hence is the key parameter for the performance of the combined estimators.

- As binary outcomes, we generate $Y_{ij} \sim Bin(p_{ij})$, $p_{ij} = 1/(1 + \exp(\mathbf{X}_{ij}^T \boldsymbol{\beta} - p_j))$ and $p_j = \log((p_0 + j\Delta_p)/(1 - p_0 - j\Delta_p))$, with $\boldsymbol{\beta} = 0.2\mathbf{1}$ and varying $\Delta_p$. Using a smaller $\boldsymbol{\beta}$ than that for the normal distribution is to avoid extreme $p_{ij}$. With this setting, without the effect of $\mathbf{X}_{ij}$, we have $p_{i1} = p_0 + \Delta_p$ and $p_{i0} = p_0$, hence $\Delta_p$ is the difference between $p_{i1}$ and $p_{i0}$.

- Performance measures that we are interested in are the variance, MSE, and coverage of 95% CI using either an asymptotic normal distribution or percentiles based on the two combined estimators. For the normal approximation, the variances are estimated from the BB samples.

- Sample sizes are $n_0 = 50$ or $n_0 = 100$, and $n_1 = n_0$ or $n_1 = 3n_0$.

- To estimate the variance and MSE, 5000 simulations were run, each with 100 bootstraps. The calculation for the coverage of 95% CI used 300 bootstraps and 3000 simulation runs.

- The contribution of external data is capped at either 100% (C=1) or 50% (C=0.5) of internal data, regardless of $n_1$.

Figures 1 and 2 show the MSE and variance of the maxML and minMSE estimators for different $\Delta_p$ (binary outcome) and $\Delta_x$ (normal outcome) when C=1, $n_0 = 100$, $n_1 = n_0$ and $n_1 = 3n_0$, respectively, with the variance of the estimator without borrowing as a reference (horizontal line). For both estimators, their variances and MSEs are low when $\Delta_p$ or $\Delta_x$ is very small or very large, as expected, since moderate values of $\Delta_p$ or $\Delta_x$ make the worst scenarios. The maxML estimator has an advantage when $\Delta_x = 0$ or $\Delta_p = 0$, which is probably the consequence of full borrowing when $\Delta_p$ or $\Delta_x$ is small. However, in general, the minMSE estimators have a lower MSE, especially when $\Delta_p$ or $\Delta_x$ has moderate values. With small $\Delta_p$ or $\Delta_x$, the minMSE estimators have a smaller MSE than the variance of no borrowing. With $n_1 = 3n_0$, both estimators improve. In particular, when $\Delta_x = 0$, the variance of maxML is close to the full borrowing one (0.008, half of that without borrowing), while that of minMSE is slightly higher. This seems rather different from the simulation reported by [19], in which a very limited benefit of borrowing is shown. This may be due to the difference between $a^*$ and $a_n^*$, which is the same as Gaywey's.

To examine the impact of using a lower cap, the same simulation was run with $C=0.5$ and the results are shown in Figures 3 and 4. The change in $C$ has a minimum impact when $n_1 = n_0$, but when $n_1 = 3n_0$ Figures 4 show a slight loss of efficiency when $\Delta_x = 0$ or $\Delta_p = 0$, but better performance in the worst scenarios.

Figures 5 and 6 present the coverage of 95% CI based on asymptotic normality or percentiles of the BB sample using both estimators for normally distributed outcomes. When $\Delta_x = 0$ or $\Delta_p = 0$, the 95% CI based on the normal approximation has slightly higher coverage, while those based on percentiles have slightly lower coverage than the nominal level. In general, the normal approximation-based CI shows the lowest coverage, suggesting that the sample size may not be sufficient for using asymptotic normality, while the two percentile-based ones have a better coverage. The sample sizes have a large impact, especially for moderate $\Delta_p$ and $\Delta_x$, around which the coverage of all CIs is generally closer to 95% when $n_0 = 100$ than it is when $n_0 = 50$. Similar patterns are observed when $n_1 = 3n_0$, (Figure 6), with better performance than $n_1 = n_0$ for moderate $\Delta_p$ and $\Delta_x$. Setting a lower cap at $C = 0.5$ improves coverage, except when $\Delta_x = 0$ or $\Delta_p = 0$ (Figures 7 and 8).

The same simulation was performed with binary outcomes and the coverage of 95% CIs are shown in Figures 9, 10, 11 and 12. In general, similar patterns as those with normally distributed outcomes have been observed. When $n_1 = n_0$ (Figures 9), although the coverage is better for worst scenarios, the trend of improvement toward the high end of $\Delta_x$ or $\Delta_p$ are less obvious, especially when $n_0 = 50$. Interestingly, the nonparametric CI based on maxML performs better than the counterpart of the minMSE estimator. Unlike in the case of normally distributed outcomes, increasing $n_1$ from $n_1 = n_0$ to $n_1 = 3n_0$ does not improve coverage in Figures 10, but lowering the cap to $C = 0.5$ does (Figures 11 and 12).

Finally, we performed a simulation with a t-distribution with df=3 for the error term for the normal distribution for the scenario of $n_1 = n_0$ only (Figures 13). Not much impact has been observed when $n_0 = 100$, but when $n_0 = 50$, the coverage rates did not converge to the nominal level at the high end of $\Delta_x$ or $\Delta_p$.

# 7 An illustrative example

We illustrate our approach using two publicly available datasets that evaluate gemtuzumab ozogamicin (GO, a CD-33 targeted therapy) with chemotherapy for children and adolescents with AML: AML03P1 [28] (Cooper et al., 2012) and AML0531 [29] (Gamis et al., 2014). Both trials had a GO with the chemotherapy arm that consisted of a remission induction phase (course 1) followed by an intensification phase (course 2). We consider complete remission (CR) at the end of course 2, a binary indicator, as the outcome of interest. There are 59 and 234 patients who had the outcome status ascertained from trial AML03P1 and AML0531 trials, respectively.

Our aim is to borrow data from AML0531 to strengthen the small arm in AML03P1. We use IPW to control the difference in the following baseline factors: log-age, log-bone marrow leukemic blast percentage (log-BM), central nervous system (CNS) disease, race, risk group and white blood count (WBC) at diagnosis. Table 2 presents the fitted PS model together with the mean / percentage differences of the above covariates. There is a substantial difference in log-WBC and also smaller differences in race and risk group. The weighted differences are generally much smaller, in particular in log-WBC, showing the effect of covariate balancing with IPW. To apply the proposed approach, the algorithm presented in Section 3 is implemented in R. Figure 14 shows the posterior density, median, and 95% credible interval of the CR rate in AML03P1 with Bayesian dynamic borrowing with and without IPW adjustment, compared to those with full and no borrowing. The density with full borrowing is rather different from the original AML03P1 one (no borrowing), while the density with dynamic borrowing is in between the two. The IPW-adjusted one is more similar to the original one, but with less variability. Table 1 gives the posterior mean, median, SE and $\sqrt{MSE}$ of CR rate with four methods that combine the two AML studies, compared to no-borrowing and full-borrowing. These results show, when the difference between the trials is significant, the advantage of IPW adjustment before applying the Bayesian dynamic borrowing with both maxML and minMSE approaches. The minMSE approach has a lower MSE than that of maxML with and without IPW adjustment.

# 8 Discussion

This work serves multiple purposes, but focus on minMSE rules as an alternative to the likelihood function based methods such as the empirical Bayesian rule, for borrowing external controls. In addition to the cminMSE rule, we propose a simplified version which may have better performance in some practical scenarios. For statistical inference, we also proposed an approach to Bayesian inference with dynamic borrowing including determining the amount of borrowing and possibly pre-adjustment for covariate differences. One advantage of the minMSE approach is the property of minimum MSE, which holds for small sample sizes as well, as it does not depend on asymptotic normality, as we have shown in the simulation. This property has been shown in almost all of our simulation results and the difference is often considerably large. Our BB based approach together with the minMSE rule is simple and can be implemented easily, e.g., in R. Our codes for the simulation in the appendix can be adapted for new simulation or other applications.

For pre-adjustment for covariates, we only mentioned IPW adjustment in our approach for simplicity. Nevertheless, our approaches can also use, together with other adjustments, for example, the doubly robust (DR) adjustment. The use of BB approach with DR has been proposed by Graham et al.[30], therefore posterior samples of $\mu_c$ with DR adjustment can be easily obtained. Wang et al. [13] detailed this approach for dynamic borrowing-based maxML. The algorithm we proposed here is similar to that in theirs, with the major difference in the rule of determining $a^*$.

To measure the performance of each approach, we have mainly focused on the MSE rather than the coverage of CI, although our simulation study shows that dynamic borrowing affects the coverage of 95% CI, due to the variability it introduced. This occurs in both the maxML and minMSE results. But the impact is generally small on quantile-based CI when the mean outcomes are similar in two data sources and when $\hat{a}^*$ is capped. These results indicate the importance of pre-adjusting the external outcome with key prognostic factors and setting a low cap for $\hat{a}^*$. A practical implication is to lower the expectation of how much can be borrowed. Our experience is that the contribution of external data should be a small fraction of the internal data, which is still beneficial.

Some previous work casts a critical view on dynamic borrowing over the difficulty of controlling type I error, including Galwey [19], Kopp-Schneider et al. [31]. Indeed, without any assumption on external data, power gains are typically not possible if strict type I error control is needed. However, one can calibrate the CI or rejection boundary to achieve it, e.g., as did Nikolakopoulos et al. [32] in the normal outcome setting. To our best knowledge, most dynamic borrowing approach did not use re-calibration, hence further investigation is needed in this direction.

Table 1: Posterior mean, median, SE and $\sqrt{MSE}$ of response rate at course 2 with four methods combining two AML studies, compared with no and full borrowing.

|             | maxML | maxML + IPW | Full bor. | No bor. | minMSE | minMSE+IPW |
|---:|---:|---:|---:|---:|---:|---:|
| Mean        | 0.942 | 0.952 | 0.909 | 0.971 | 0.963 | 0.964 |
| SE          | 0.023 | 0.016 | 0.016 | 0.024 | 0.025 | 0.019 |
| $\sqrt{MSE}$ | 0.038 | 0.025 | 0.064 | 0.024 | 0.027 | 0.020 |

Table 2: Summary of fitted logistic PS model for probability of being in study AML0531, with raw and IPW weighted covariate differences (AML0531 - AML03P1).

|             | Estimate | Std. Error | z value | Pr(>|z|) | Raw Diff. | Weighted Diff. |
|---:|---:|---:|---:|---:|---:|---:|
| CNS disease | 0.189  | 0.693 | 0.273  | 0.785 | -0.042 | -0.011 |
| Race        | -0.783 | 0.453 | -1.731 | 0.084 | 0.043  | 0.036  |
| log_MRD     | 0.059  | 0.132 | 0.449  | 0.653 | -0.044 | -0.043 |
| log_age     | -0.082 | 0.153 | -0.538 | 0.591 | 0.057  | -0.209 |
| log_WBC     | -0.282 | 0.119 | -2.371 | 0.018 | 0.553  | 0.118  |
| log_BM      | -2.703 | 1.942 | -1.392 | 0.164 | 0.242  | 0.068  |
| low_risk    | -0.187 | 0.349 | -0.535 | 0.592 | -0.035 | -0.080 |
| high_risk   | 1.199  | 0.693 | 1.728  | 0.084 | 0.048  | 0.028  |

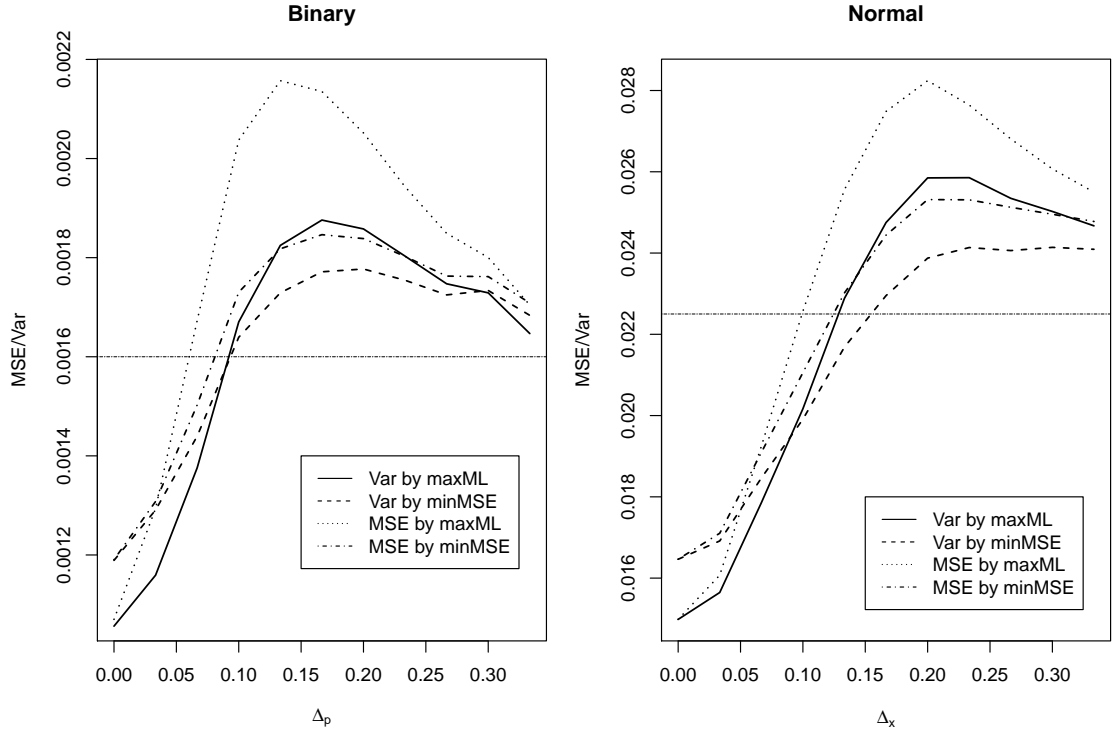

Figure 1: Simulation results to compare MSE and variance of maxML and minMSE estimators based combinations, normal and binary outcomes.

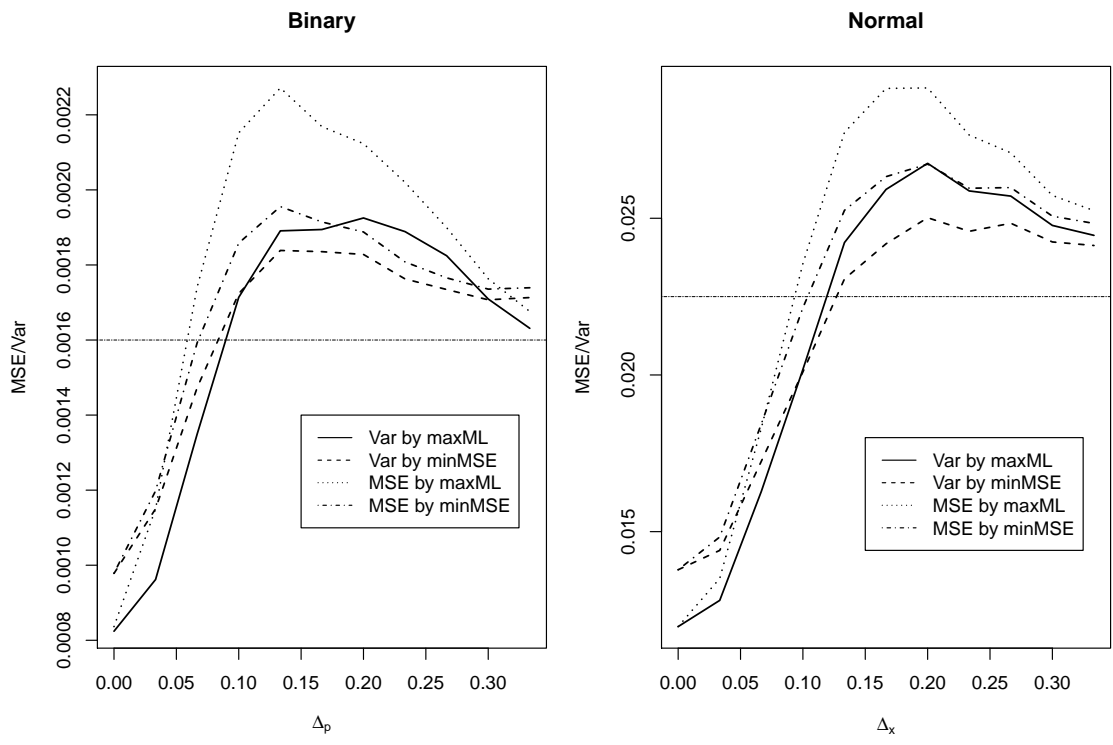

Figure 2: Simulation results to compare MSE and variance of maxML and minMSE estimators based combinations, normal and binary outcomes, with $n_1 = 3n_0$.

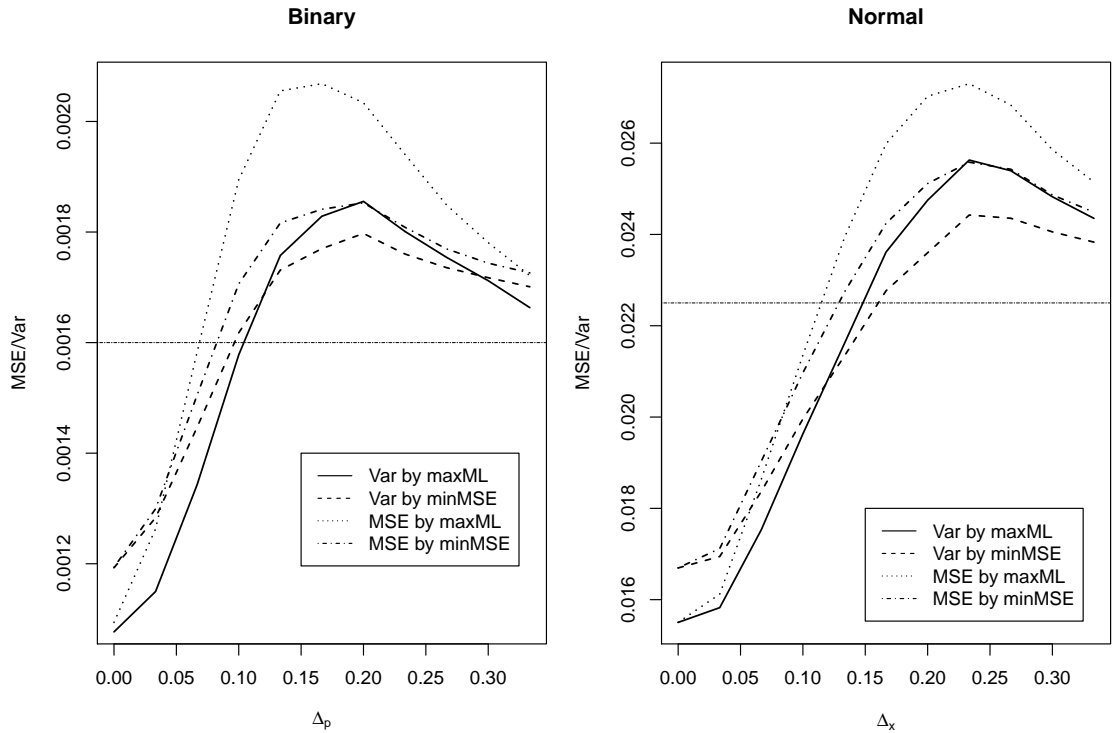

Figure 3: Simulation results to compare MSE and variance of maxML and minMSE estimators based combinations, normal and binary outcomes, capped at C=0.5.

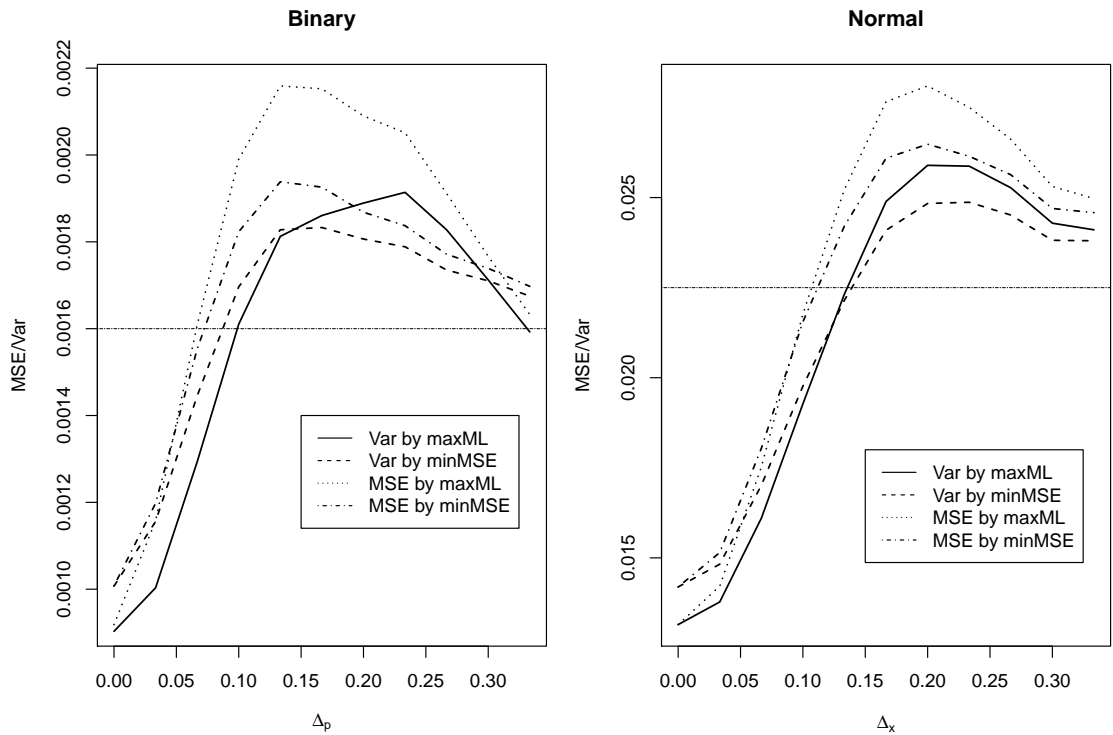

Figure 4: Simulation results to compare MSE and variance of maxML and minMSE estimators based combinations, normal and binary outcomes, with $n_1 = 3n_0$, capped at C=0.5.

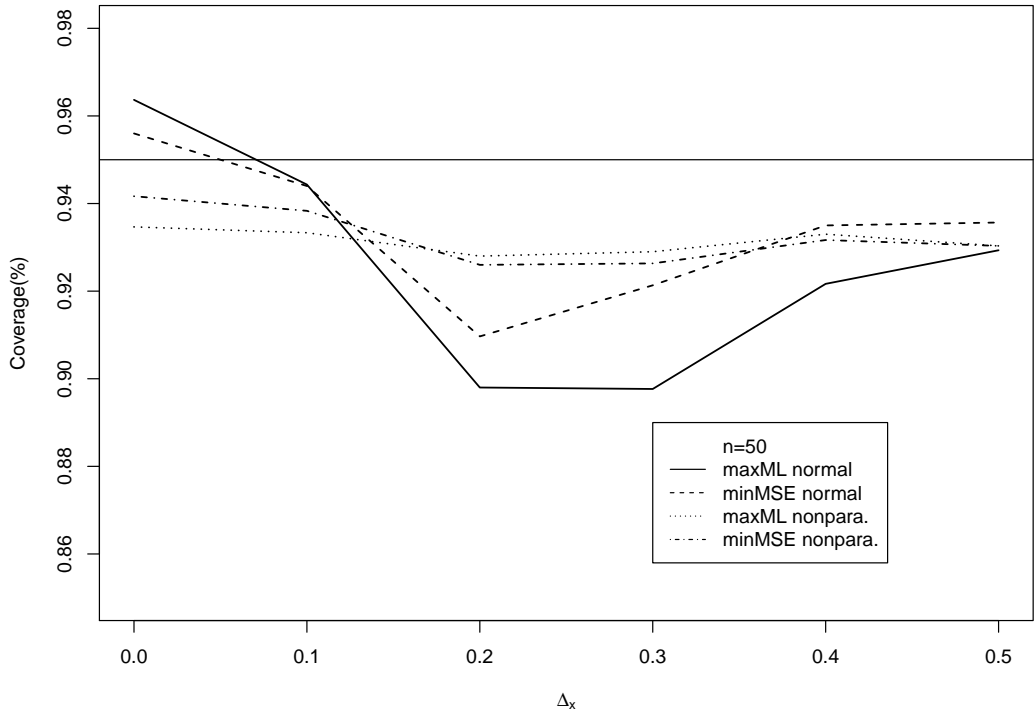

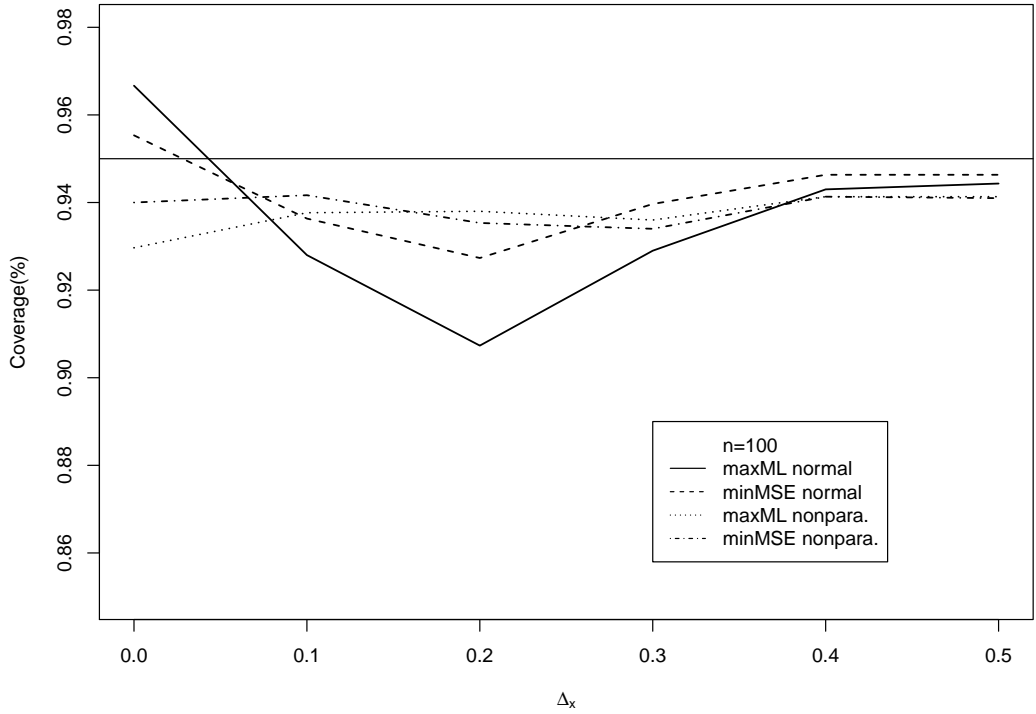

Figure 5: Simulation results comparing 95% coverage of maxML and minMSE estimators based combinations by bias, normal outcome with sample sizes 50 and 100.

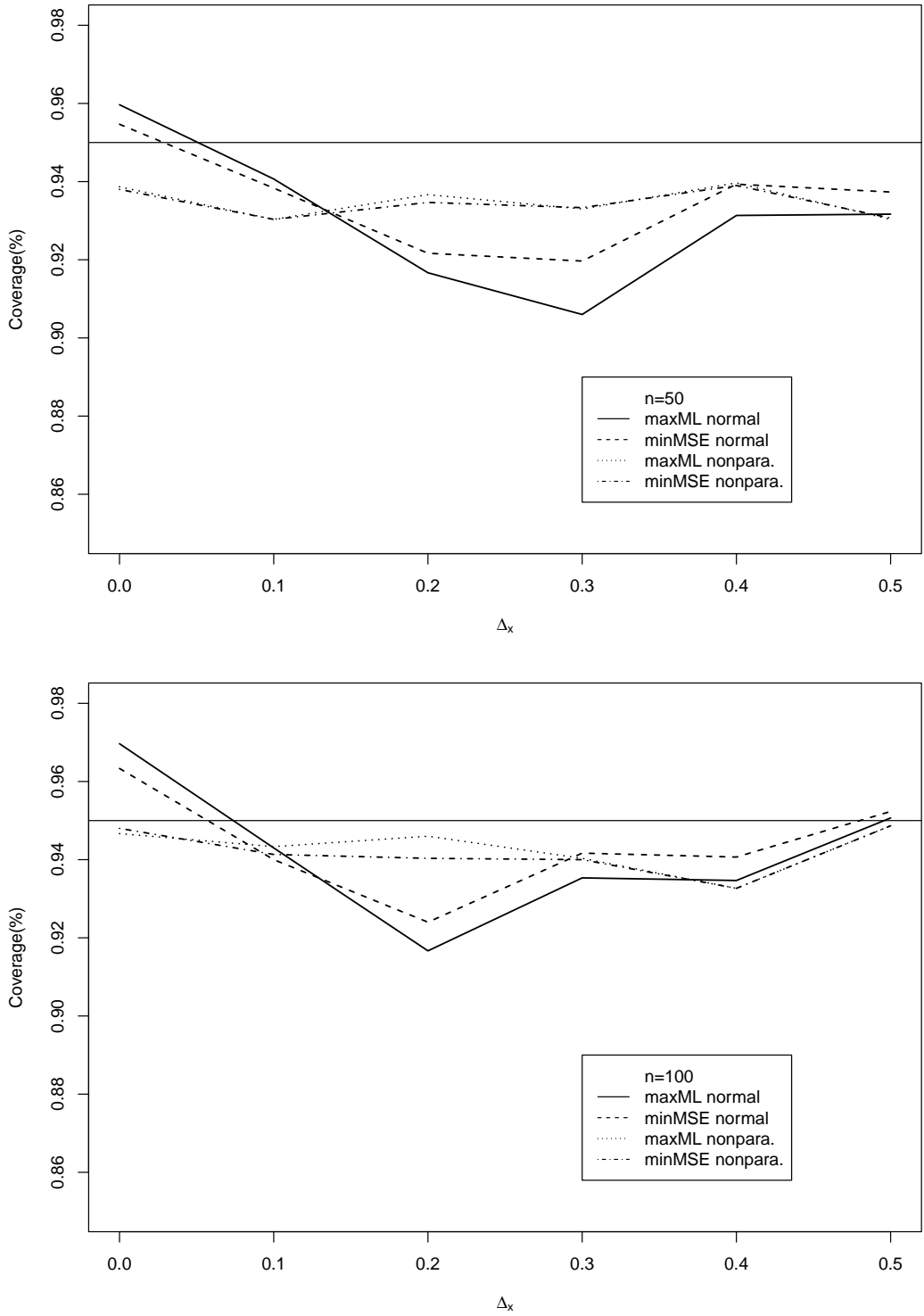

Figure 6: Simulation results comparing 95% coverage of maxML and minMSE estimators based combinations by bias, normal outcome with sample sizes $n_0 = 50$ and $100$, $n_1 = 3n_0$.

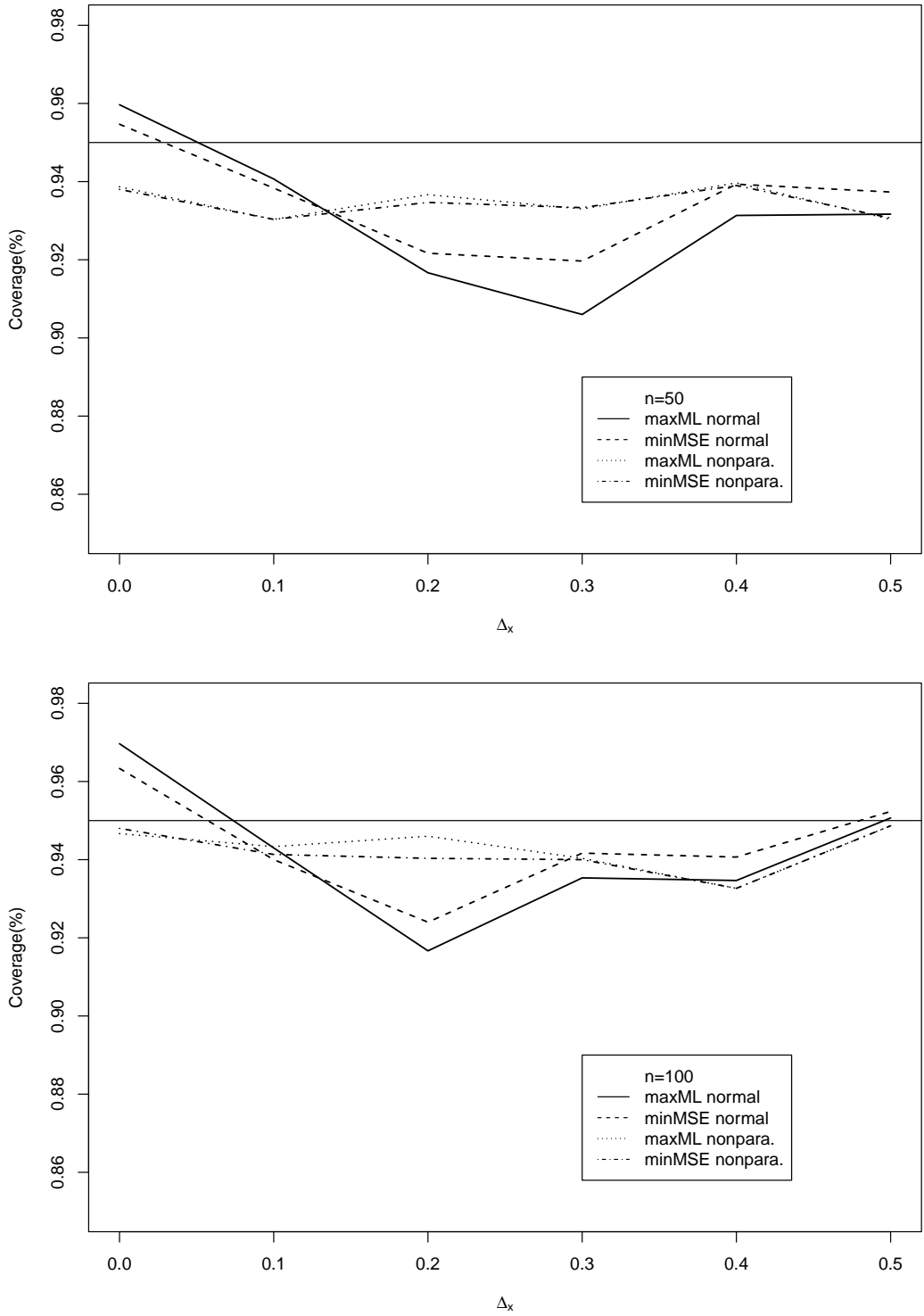

Figure 7: Simulation results comparing 95% coverage of maxML and minMSE estimators based combinations by bias, normal outcome with sample sizes 50 and 100, capped at C=0.5.

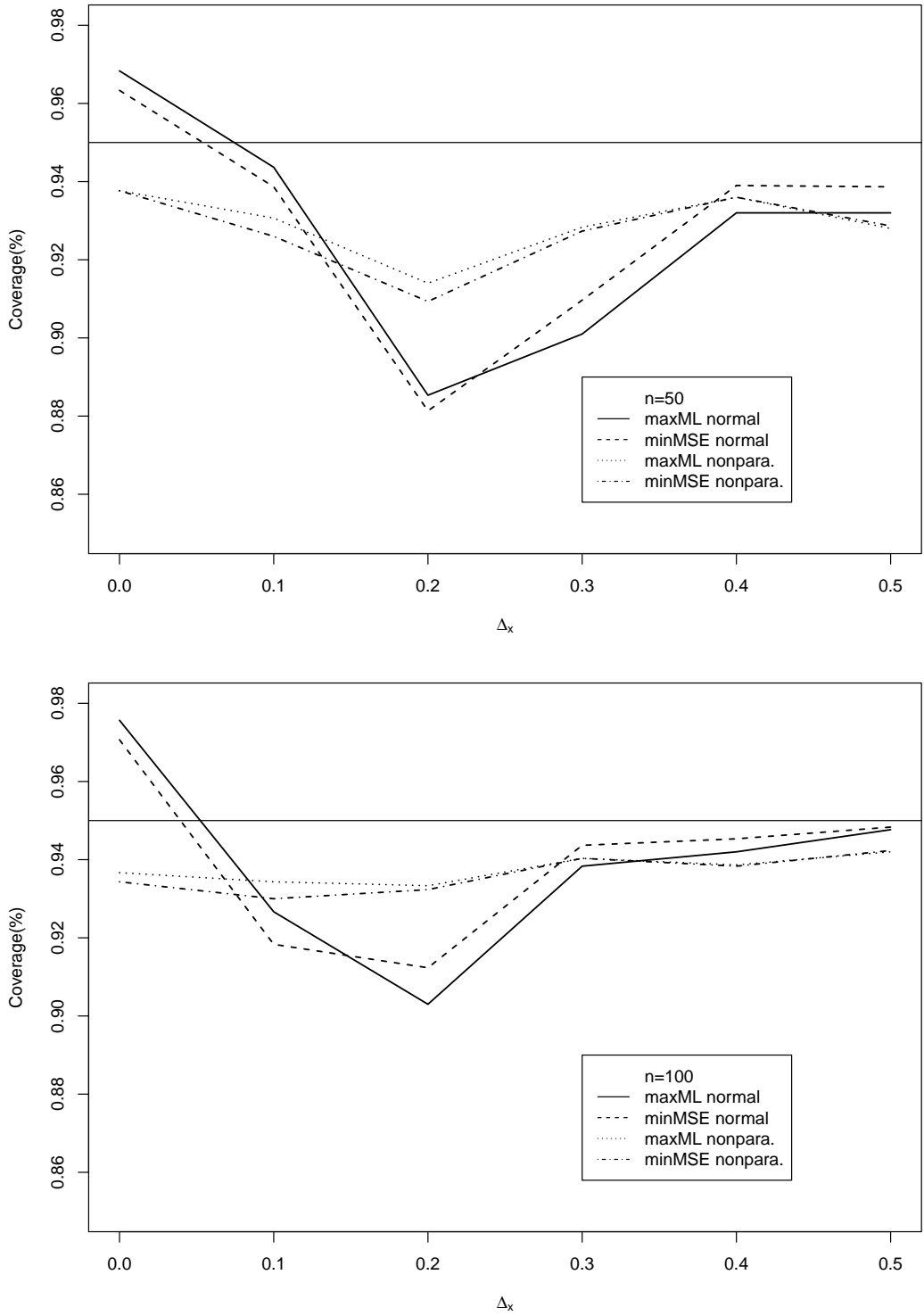

Figure 8: Simulation results comparing 95% coverage of maxML and minMSE estimators based combinations by bias, normal outcome with sample sizes $n_0 = 50$ and $100$, $n_1 = 3n_0$, capped at C=0.5.

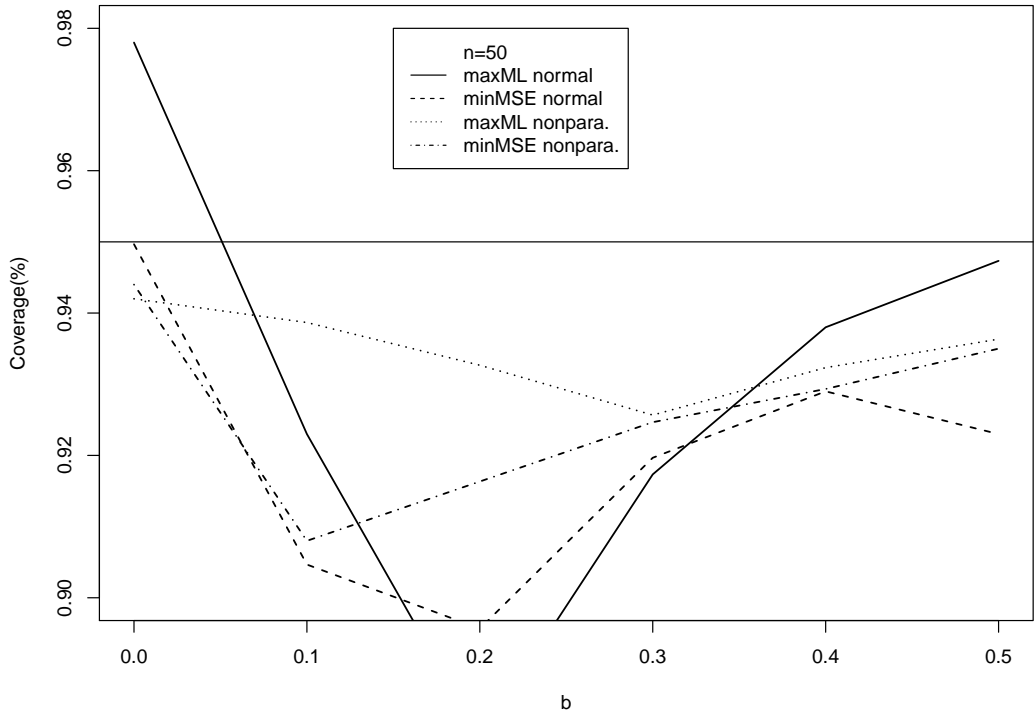
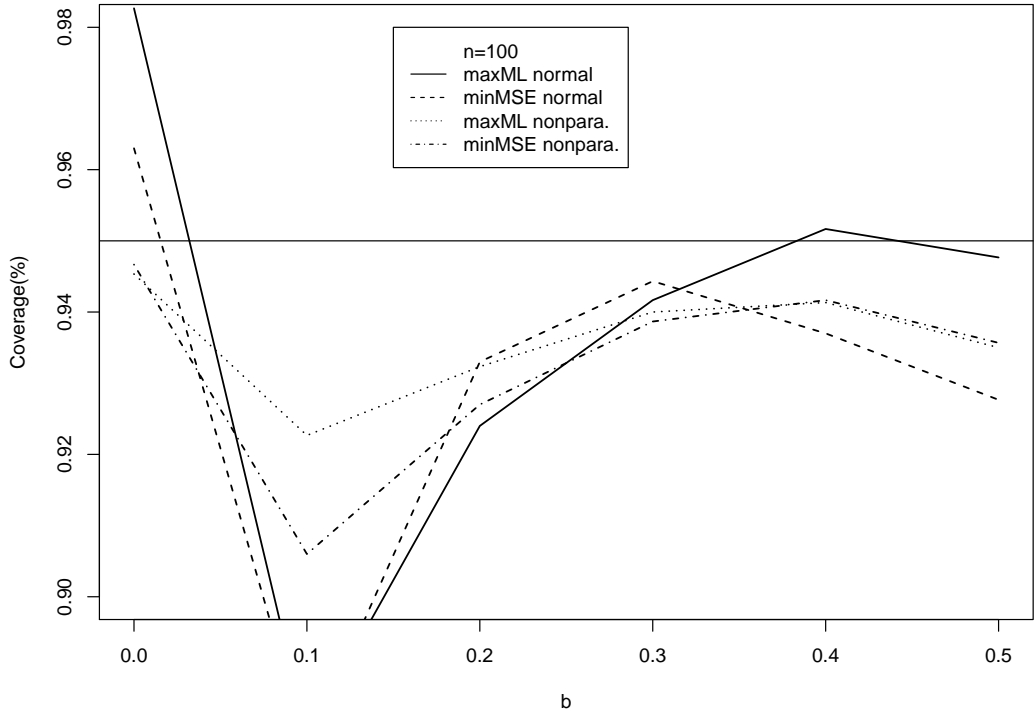

Figure 9: Simulation results comparing 95% coverage of maxML and minMSE estimators based combinations by bias, binary outcome with sample sizes 50 and 100.

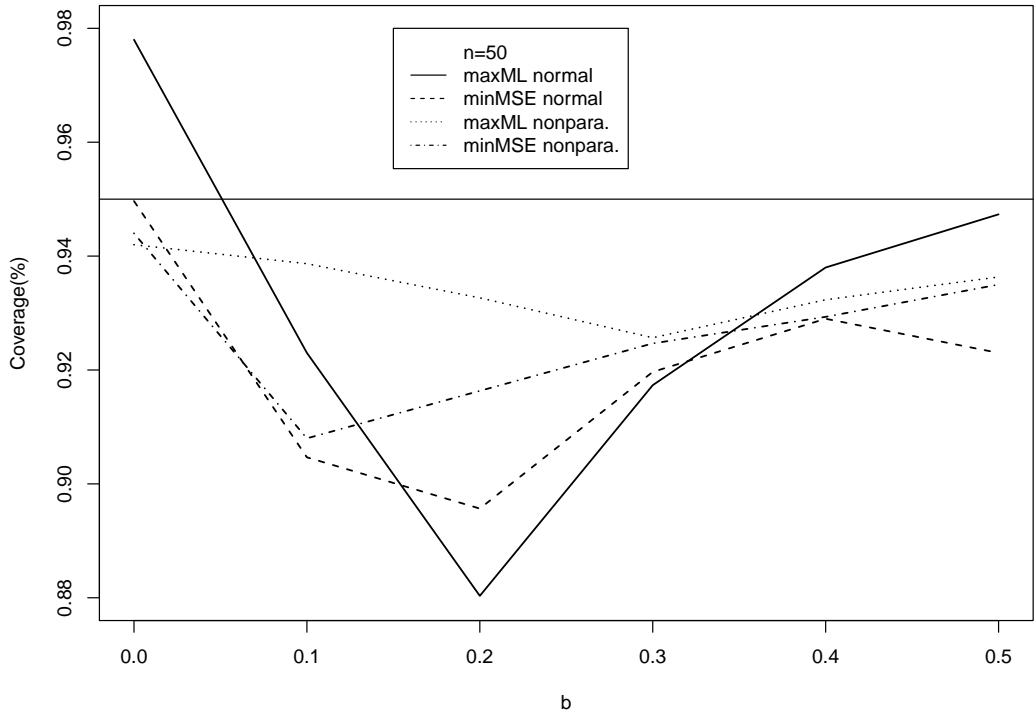

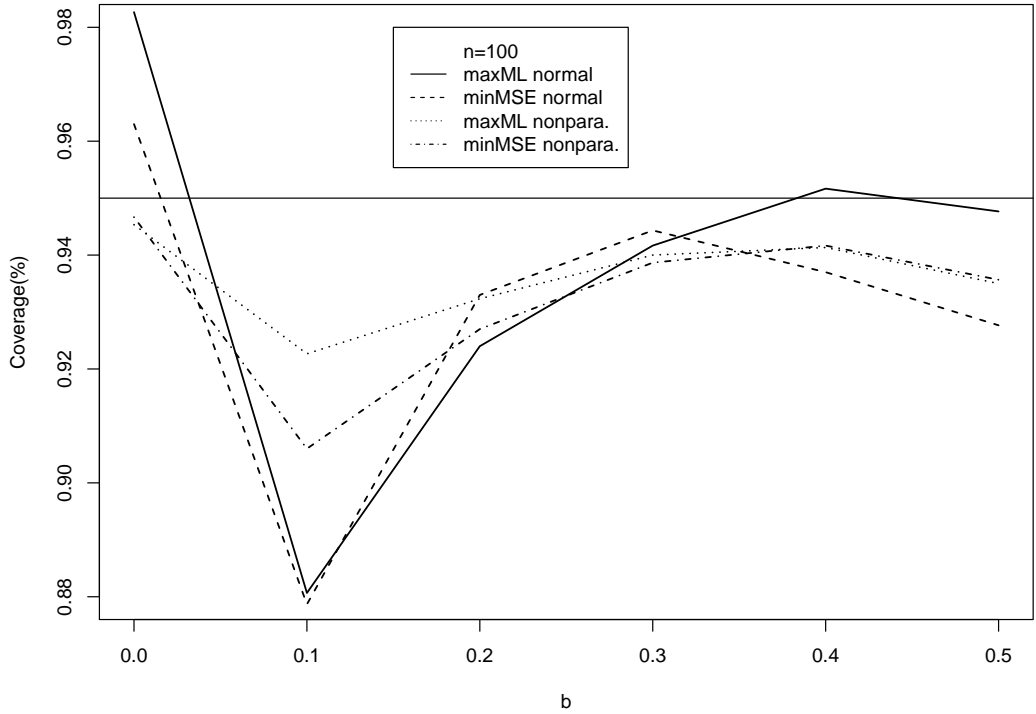

Figure 10: Simulation results comparing 95% coverage of maxML and minMSE estimators based combinations by bias, binary outcome with sample sizes 50 and 100.

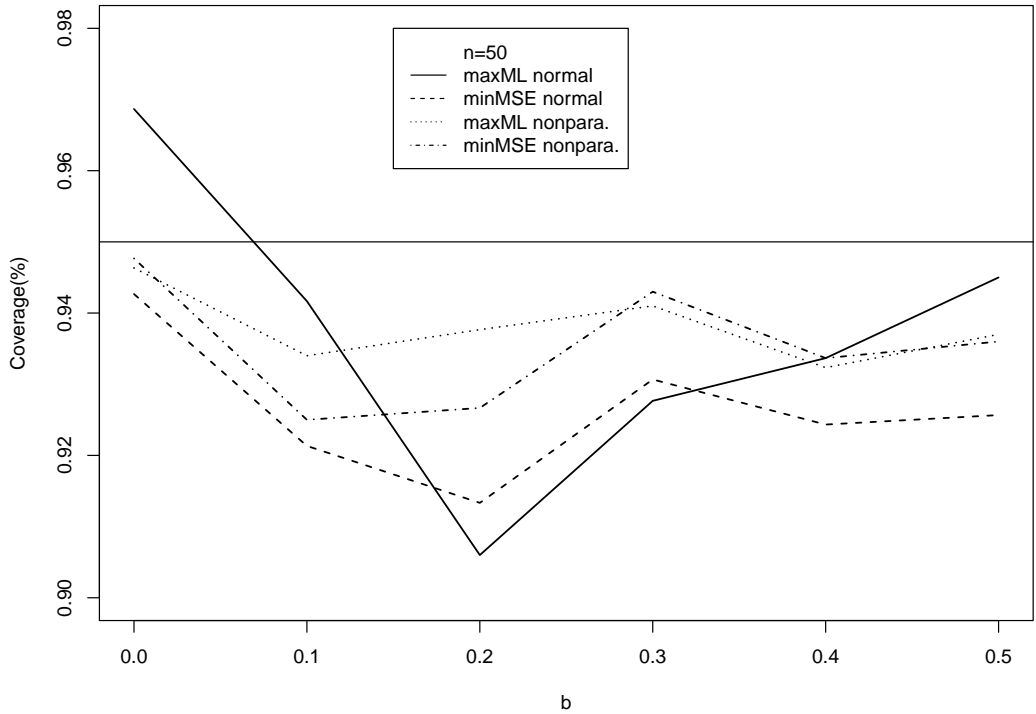

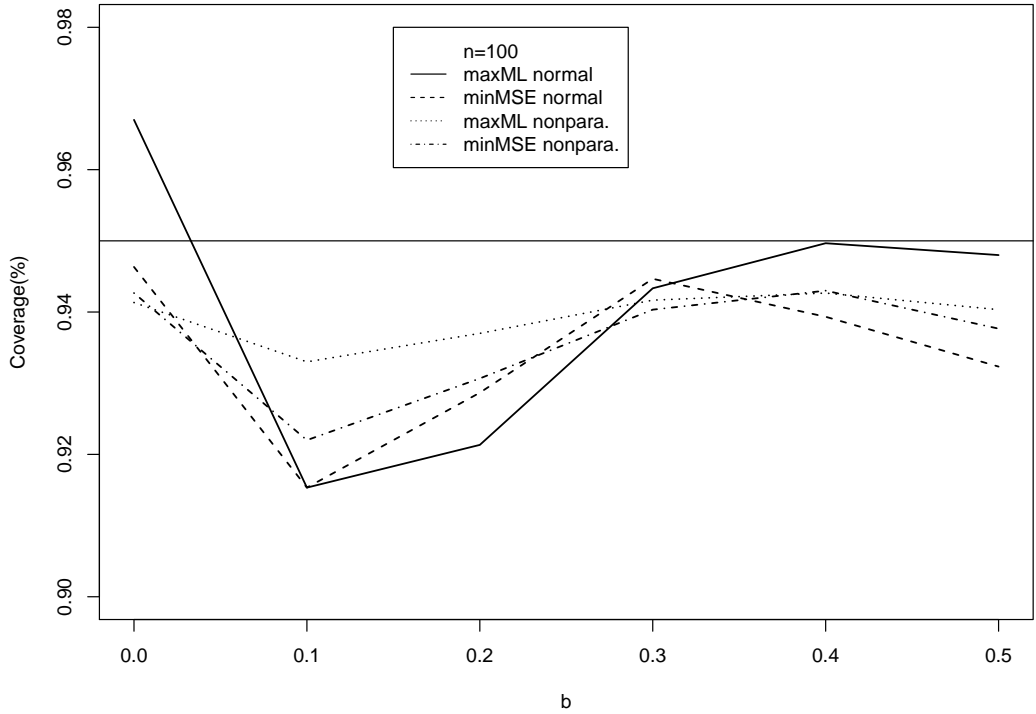

Figure 11: Simulation results comparing 95% coverage of maxML and minMSE estimators based combinations by bias, binary outcome with sample sizes 50 and 100, capped at C=0.5.

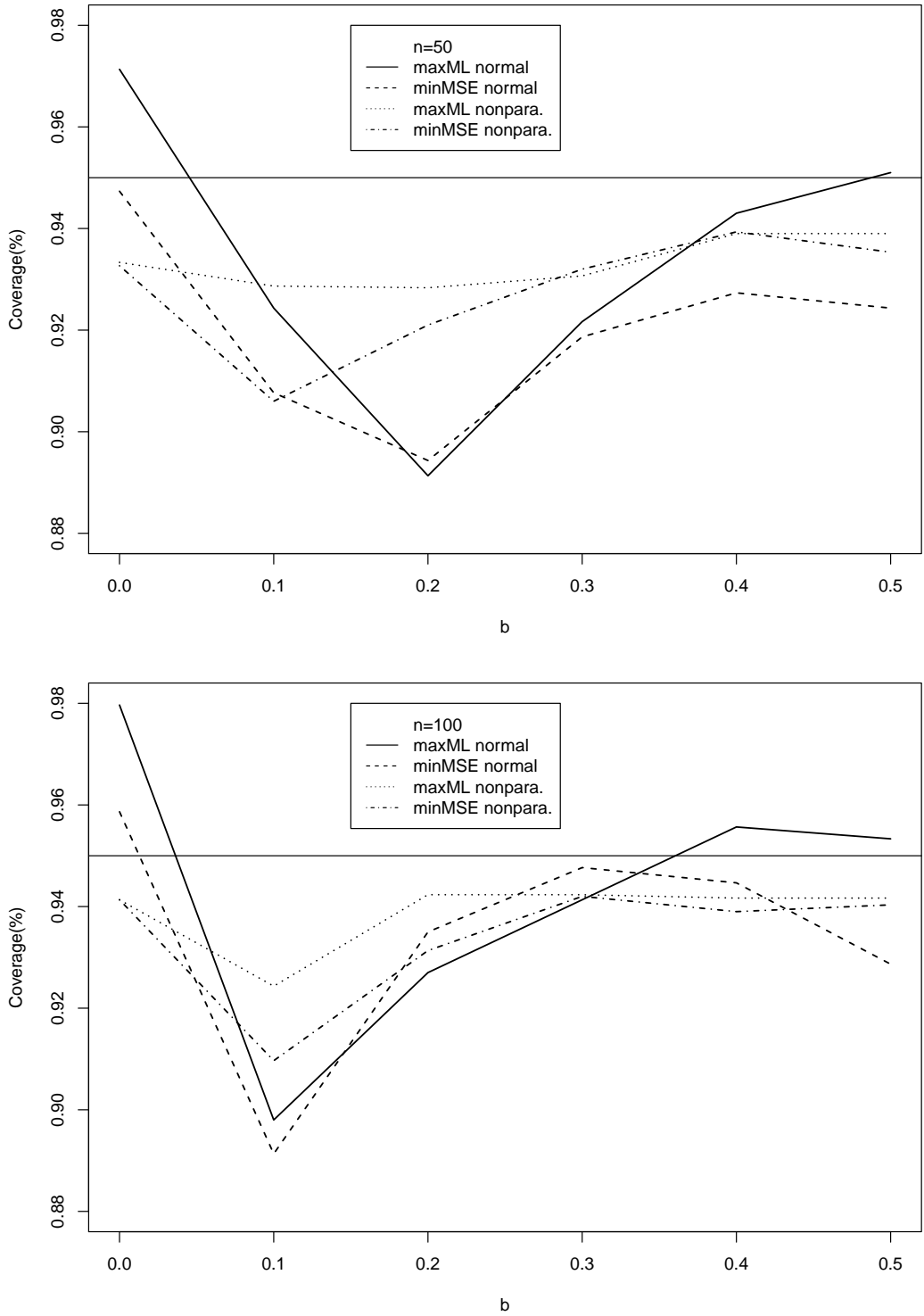

Figure 12: Simulation results comparing 95% coverage of maxML and minMSE estimators based combinations by bias, binary outcome with sample sizes 50 and 100 , capped at C=0.5.

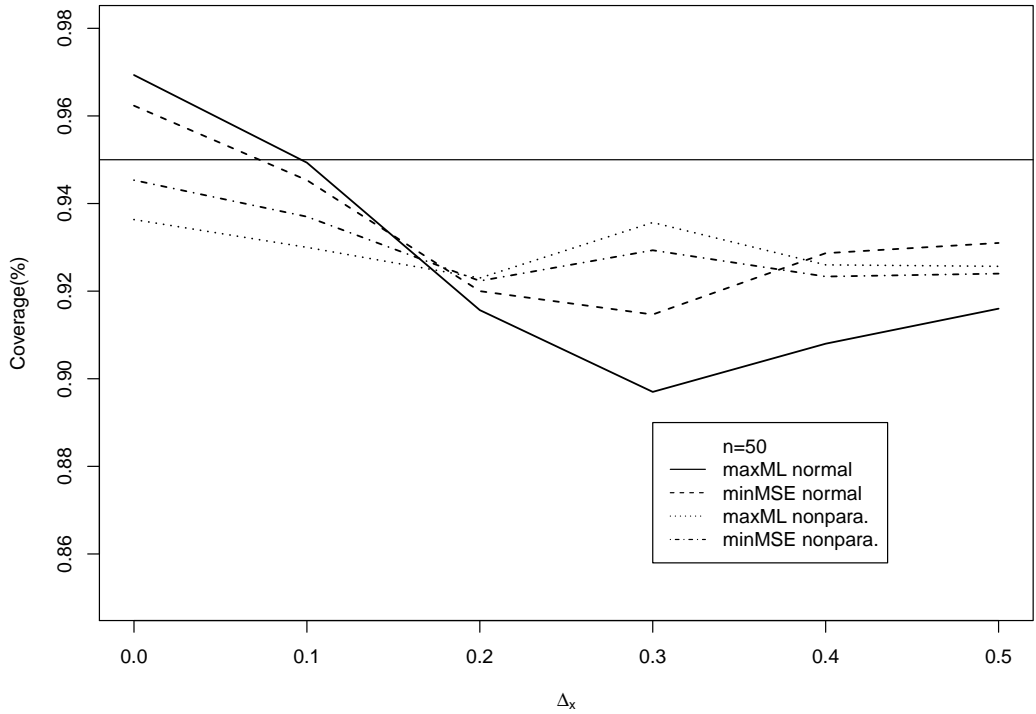

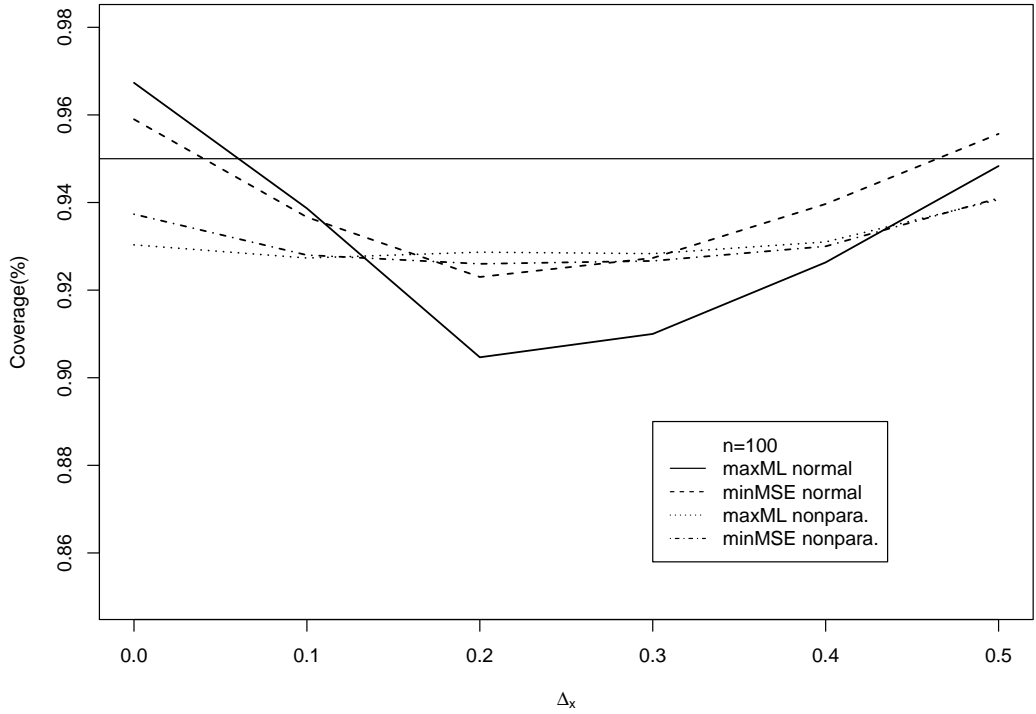

Figure 13: Simulation results comparing 95% coverage of maxML and minMSE estimators based combinations by bias, t-distributed outcome with df=3 with sample sizes 50 and 100.

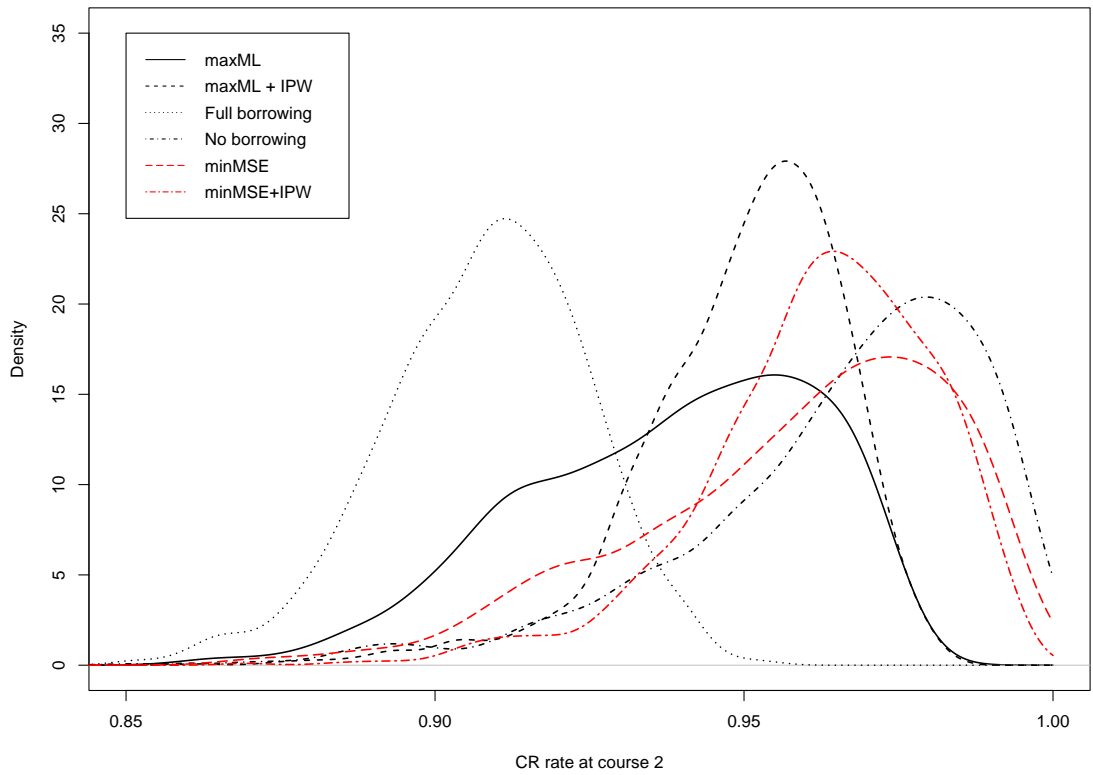

Figure 14: Distribution of CR rate at cycle 2 by MLE and minMSE combinations with and without IPW adjustment, compared with no and full borrowing

# 9 Appendix

## 9.1 Derivation of minMSE rule

Here we give details of the derivation for Section 4.1. We start from Eq (16)

$$MSE(\hat{\mu}^c(a)) = \frac{\sigma_0^2 + a^2(\sigma_1^2 + \delta^2)}{(1+a)^2}. \tag{31}$$

Setting

$$\frac{\partial MSE(\hat{\mu}^c(a))}{\partial a} \propto [(1+a)a(\sigma_1^2 + \delta^2) - \sigma_0^2 - a^2(\sigma_1^2 + \delta^2)] = 0 \tag{32}$$

yields optimal $a^*$:

$$a^* = \sigma_0^2/(\sigma_1^2 + \delta^2), \tag{33}$$

which is the same as that of Galwey (2017) in a different setting. With this $a^*$, we have the following posterior mean of $\mu_c(a^*)$

$$\mu_c(a^*) = \frac{\mu_0 + a^*\mu_1}{1 + a^*}. \tag{34}$$

The bias of $\hat{\mu}^c(^*a)$ is

$$\Delta(a^*) = \mu_c(a^*) - \mu_0 \tag{35}$$

$$= \frac{a^*\delta}{1 + a^*} \tag{36}$$

$$= \frac{\delta\sigma_0^2}{\sigma_1^2 + \sigma_0^2 + \delta^2} \tag{37}$$

which tends to zero even when $\delta \to \infty$. The maximum squared bias can be found by letting

$$\frac{\partial \Delta^2(a^*)}{\partial \delta^2} \propto [\sigma_1^2 + \sigma_0^2 + \delta^2 - 2\delta^2] = 0, \tag{38}$$

which leads to $\delta^2 = \sigma_1^2 + \sigma_0^2$, that is, the maximum $\Delta^2(a^*)$ occurs when $\delta^2$ equals the total variance.

## 9.2 Asymptotic properties

We derive asymptotic properties of the posterior sample $\hat{\mu}_c^b(\hat{a}^*)$ based on the results in [20, 21] when $n_j \to \infty, j = 0, 1$. Without loss of generality, we assume that $n_1 = n_0 = n$. First, we consider the asymptotic properties of $\hat{\mu}_c^b(a)$ with fixed $a$. Since

$$\sqrt{n}(\hat{\mu}_c^b(a) - \hat{\mu}_c(a)) = (1+a)^{-1}[\sqrt{n}(\hat{\mu}_0^b - \hat{\mu}_0) + a\sqrt{n}(\hat{\mu}_1^b - \hat{\mu}_1)] \sim_a (1+a)^{-1}(Z_0 + aZ_1) \tag{39}$$

where $Z_k \sim N(0, \sigma_k^2)$ only depends on data $D_k$, based on Theorem 1, Ref [20], and $C_n \sim_a C$ means $C_n$ weakly converges to $C$. Because $\hat{a}^* \to a^*$, a.s., using the continuous mapping theorem (Section 2.7, [33], we have

$$\sqrt{n}(\hat{\mu}_c^b(\hat{a}^{*b}) - \hat{\mu}_c(\hat{a}^*)) \sim_a (1+a^*)^{-1}(Z_0 + a^*Z_1) \tag{40}$$

Theorem 1 in Ref [20] also states that $\mu_c(a^*)$, given $D_0, D_1$ has the same asymptotic posterior distribution as $\hat{\mu}_c^b(\hat{a}^{*b})$, that is,

$$\sqrt{n}(\mu_c(a^*) - \hat{\mu}_c(\hat{a}^*)) \sim_a \sqrt{n}(\hat{\mu}_c^b(\hat{a}^{*b}) - \hat{\mu}_c(\hat{a}^*)) \tag{41}$$

Therefore, inference based on $\hat{\mu}_c^b(\hat{a}^{*b})$ is indeed an approximate Bayesian inference.

Note that when $\mu_1 \neq \mu_0$, $\hat{a}^* \to 0$ a.s. and we have

$$\sqrt{n}(\hat{\mu}_c^b(\hat{a}^{*b}) - \hat{\mu}_c(\hat{a}^*)) \sim_a Z_0 \tag{42}$$

which is independent of $D_1$. Therefore, a test, eg., for $H_0 : \mu_c = 0$ based on $\hat{\mu}_c^b(\hat{a}^{*b})$ asymptotically controls the type I error. This is a property of the elastic priors [25]. Although $\hat{a}^*$ only has half of the information borrowing consistent property, the key request of type I error control still holds. Although asymptotic distributions can be derived, using the percentiles of BB posterior samples for CI may be a better choice than relying on asymptotic normality. As shown by the simulation results, the former has better coverage especially with moderate $\Delta_x$ values.

### 9.3 R-code

The following codes produces Figure 1

```
--------------The following code produces Figure 1
library(mvtnorm)
simu=function(
  p=5,
  nc=100,
  nh=3*nc,
  dx=0, cap=1,
  nsimu=50000
)
{
  Out=NULL
  for(BB in 1:nsimu){
    Xi=rmvnorm(nc,rep(0,p),diag(1,p))
    Xh=rmvnorm(nh,rep(0,p),diag(1,p))+dx
    bbeta=rep(0.5,p)
    yc=Xi %*% bbeta+rnorm(nc)
    yh=Xh %*% bbeta+rnorm(nh)
    sigc=var(yc)/nc
    sigh=var(yh)/nh
    muc=mean(yc)
    muh=mean(yh)
    del=muh-muc
    a=min(cap,sigc/(sigh+del^2))
    mub=(muc+a*muh)/(1+a)
    a0=min(cap*nc/nh,sigh/(max((muc-muh)^2,sigc+sigh)-sigc))
    sig0=1/(1/sigc+ a0/sigh)
    mue=(muc/sigc+a0*muh/sigh)*sig0
    Out=rbind(Out,c(mue,mub))
  }
  mm=apply(Out,2,mean)
  mvar=apply(Out,2,var)
  c(nc,dx,mvar,mvar+mm^2)
}

simub=function(
  p=5,
  nc=100,
```

```
    nh=3*nc,
    dx=0,bb=0.0,
    p0=0.2, cap=1,
    nsimu=50000
)
{
  Out=NULL
  for(BB in 1:nsimu){
    yc=rbinom(nc,1,p0)
    yh=rbinom(nh,1,p0+dx)
    muc=sum(yc)
    muh=sum(yh)
    va0=(0:50)/50
    ll=lbeta(va0*muh+muc+1,va0*(nh-muh)+nc-muc+1)-lbeta(va0*muh+1,va0*(nh-muh)+1)
    a00=min(cap*nc/nh,max(va0[ll==max(ll)]))
    mue=(a00*muh+muc+1)/(nc+a00*nh+2)
    sigc=(muc+1)*(nc-muc+1)/(nc+2)^2/(nc+3)
    sigh=(muh+1)*(nh-muh+1)/(nh+2)^2/(nh+3)
    del=muh/nh-muc/nc
    a=min(cap,sigc/(sigh+del^2))
    muc=(mean(yc)+a*mean(yh))/(1+a)
    Out=rbind(Out,c(mue,muc)-p0)
  }
  mm=apply(Out,2,mean)
  mvar=apply(Out,2,var)
  c(nc,dx,mvar,mvar+mm^2)
}

vdx=(0:10)/30
Pout=simub(dx=vdx[1])
Pout=rbind(Pout,simub(dx=vdx[2]))
Pout=rbind(Pout,simub(dx=vdx[3]))
Pout=rbind(Pout,simub(dx=vdx[4]))
Pout=rbind(Pout,simub(dx=vdx[5]))
Pout=rbind(Pout,simub(dx=vdx[6]))
Pout=rbind(Pout,simub(dx=vdx[7]))
Pout=rbind(Pout,simub(dx=vdx[8]))
Pout=rbind(Pout,simub(dx=vdx[9]))
Pout=rbind(Pout,simub(dx=vdx[10]))
Pout=rbind(Pout,simub(dx=vdx[11]))

Rout=Pout
vdx=(0:10)/30
Pout=simu(dx=vdx[1])
Pout=rbind(Pout,simu(dx=vdx[2]))
Pout=rbind(Pout,simu(dx=vdx[3]))
Pout=rbind(Pout,simu(dx=vdx[4]))
Pout=rbind(Pout,simu(dx=vdx[5]))
Pout=rbind(Pout,simu(dx=vdx[6]))
Pout=rbind(Pout,simu(dx=vdx[7]))
Pout=rbind(Pout,simu(dx=vdx[8]))
Pout=rbind(Pout,simu(dx=vdx[9]))
Pout=rbind(Pout,simu(dx=vdx[10]))
Pout=rbind(Pout,simu(dx=vdx[11]))

pdf(file="MSE3cap.pdf",width=10, height=7)
```

```r
#--------------------The following code produces Figure 5
simuBB=function(
  p=5,
  nboot=500,
  nc=100,
  nh=nc,
  dx=0, cap=0.5,
  nsimu=3000
)
{
  qout=NULL
  for(s in 1:nsimu){
    Xi=rmvnorm(nc,rep(0,p),diag(1,p))
    Xh=rmvnorm(nh,rep(0,p),diag(1,p))+dx
    bbeta=rep(0.5,p)
    yc=Xi %*% bbeta+rnorm(nc)
    yh=Xh %*% bbeta+rnorm(nh)
    muc=mean(yc)
    muh=mean(yh)
    sigc=var(yc)/nc
    sigh=var(yh)/nh
    del=muh-muc
    a=min(cap,sigc/(sigh+del^2))
    mu=(muc+a*muh)/(1+a)
    a0=min(cap*nc/nh,sigh/(max((muc-muh)^2,sigc+sigh)-sigc))
    sig0=1/(1/sigc+ a0/sigh)
    mue=(muc/sigc+a0*muh/sigh)*sig0
    mm=c(mue,mu)
    Out=NULL
    for (BB in 1:nboot){
      wi=rexp(nc)
      vi=rexp(nh)
      wi=wi/mean(wi)
      vi=vi/mean(vi)
      muc=mean(yc*wi)
      muh=mean(yh*vi)
      sigc=wtd.var(yc,wi)/nc
      sigh=wtd.var(yh,vi)/nh
      del=muh-muc
      a=min(cap,sigc/(sigh+del^2))
      mumse=(muc+a*muh)/(1+a)
      a0=min(cap*nc/nh,sigh/(max((muc-muh)^2,sigc+sigh)-sigc))
      sig0=1/(1/sigc+ a0/sigh)
      muml=(muc/sigc+a0*muh/sigh)*sig0
      Out=rbind(Out,c(mumse,muml))
    }
    mvar=apply(Out,2,var)
    coverc=sign(mm+1.96*sqrt(mvar))*sign(mm-1.96*sqrt(mvar))<0
    quan=apply(Out,2,quantile,probs=c(0.025,0.975))
    coverm=sign(quan[1,])*sign(quan[2,])<0
    #    browser()
    qout=rbind(qout,c(coverc,coverm))
  }
  c(nc,dx,apply(qout,2,mean))
}
```

```
Pout=simuBB(dx=0)
Pout=rbind(Pout,simuBB(dx=0.1))
Pout=rbind(Pout,simuBB(dx=0.2))
Pout=rbind(Pout,simuBB(dx=0.3))
Pout=rbind(Pout,simuBB(dx=0.4))
Pout=rbind(Pout,simuBB(dx=0.5))

Qout=Pout

Pout=simuBB(nc=50,dx=0)
Pout=rbind(Pout,simuBB(nc=50,dx=0.1))
Pout=rbind(Pout,simuBB(nc=50,dx=0.2))
Pout=rbind(Pout,simuBB(nc=50,dx=0.3))
Pout=rbind(Pout,simuBB(nc=50,dx=0.4))
Pout=rbind(Pout,simuBB(nc=50,dx=0.5))

vdx=(0:5)/10
ymax=range(Pout[,3:6])

pdf(file="Covercap5.pdf",width=9, height=13)
par(mfrow=c(2,1))
par(mar=c(4.1, 4.1, 2.1, 1))
plot(x=vdx,y=Pout[,3],ylim=c(0.85,.98),xlab=expression(Delta[x]),
            ylab="Coverage(%)",lwd=1.5,type="l")
lines(x=vdx,y=Pout[,4],lty=2,lwd=1.5)
lines(x=vdx,y=Pout[,5],lty=3,lwd=1.5)
lines(x=vdx,y=Pout[,6],lty=4,lwd=1.5)
abline(h=0.95)
legend(x=0.3,y=0.89,legend=c("n=50","maxML normal", "minMSE normal"
   ,"maxML nonpara.", "minMSE nonpara."),lty=0:4)

plot(x=vdx,y=Qout[,3],ylim=c(0.85,.98),xlab=expression(Delta[x]),
       ylab="Coverage(%)",lwd=1.5,type="l")
lines(x=vdx,y=Qout[,4],lty=2,lwd=1.5)
lines(x=vdx,y=Qout[,5],lty=3,lwd=1.5)
lines(x=vdx,y=Qout[,6],lty=4,lwd=1.5)
abline(h=0.95)
legend(x=0.3,y=0.89,legend=c("n=100","maxML normal",
   "minMSE normal","maxML nonpara.", "minMSE nonpara."), lty=0:4)
dev.off()
```